# Understanding Human-Chatbot Romance:
# A Qualitative and Quantitative Study on Romantic Fantasy and Other Interpersonal Characteristics

Paula Ebner, Dr. Jessica Szczuka

**Author Note**

Paula Ebner[a], paula.ebner@uni-due.de

Dr. Jessica Szczuka[a], jessica.sczcuka@uni-due.de

[a] University Duisburg-Essen, TECTRUM, Bismarckstreet 120, 47075 Duisburg

Data, headlines, and additional online materials are openly available at the project's Open Science Framework page (https://osf.io/fs5pq/?view_only=796a730686504b5a9475b3cba430b055). We have no conflicts of interest to disclose.

Correspondence concerning this article should be addressed to paula.ebner@uni-due.de.

**Abstract**

LLM-based chatbots are now being specifically designed to facilitate social companionship, even romantic relationships, incorporating features that parallel human relationship dynamics. This has led a subset of users to form romantic relationships with chatbots. Understanding which interpersonal characteristics drive individuals to form intense, emotional bonds with chatbots is crucial for comprehending the potential psychological and societal impacts of romantic human-chatbot relationships. This mixed-methods study investigates psychological predictors of relationship intensity among individuals currently in romantic relationships with chatbots. Romantic and sexual fantasy, promising constructs not previously investiagted in this context, are examined alongside previously discussed factors (loneliness, anthropomorphism, attachment orientation, and sexual sensation seeking). In Study 1, quantitative data from individuals with chatbot partners (N=92) showed that romantic fantasy explained the most variance in relationship intensity, with additional contributions from anthropomorphism and avoidant attachment. Contrary to expectations, the other predictors, including loneliness, did not significantly predict intensity. In Study 2, 15 qualitative interviews illuminated how users employ romantic fantasy to enhance their relationships, describing active fantasy use to shape interactions and a desire for their chatbot to feel as human as possible. This study provides the first quantitative sample of this under-researched population, explaining who might form more intense romantic relationships with chatbots.

**Understanding Human-Chatbot Romance:**

**A Qualitative and Quantitative Study on Romantic Fantasy and Other Interpersonal Characteristics**

Recent advancements in text-generative AI have produced social chatbots capable of romantic and sexual interactions so sophisticated that users often cannot distinguish them from human conversation partners, enabling these systems to elicit genuine emotional responses and attachment (Skjuve et al., 2021; Kim et al., 2023; Starke et al., 2024). Some users are spending hours daily chatting with AI companions, sharing intimate thoughts, seeking emotional support, and even describing themselves as being 'in a relationship' with their chatbot. These romantic interactions with AI are becoming increasingly common, with 16% of adult singles in the United States having already experimented with AI as a romantic companion (Lee, 2025), and 16 of 100 most visited AI-based apps are designed for companionship (Malfacini, 2025). Despite these high the emergence of romantic human-chatbot relationships represents a fundamentally new intersection of advanced technology with one of humanity's most intimate social structures.

Understanding which psychological factors predict relationship intensity helps clarify who could be especially prone to experiencing strong emotional bonds and how these bonds are shaped by users' own internal processes, such as romantic and sexual fantasy, anthropomorphism, attachment tendencies, loneliness, or sensation seeking. By identifying the psychological dispositions linked to higher relationship intensity, we aim to better understand which user groups may be more likely to experience AI companions as emotionally significant. Furthermore, the findings could enable more data-driven discussions about romantic AI companionship and contribute meaningfully to debates about what makes human social interaction unique and how machines with implemented social capabilities should be situated within that landscape.

This study proposes that romantic and sexual fantasies may play a particularly important role in this context, as AI chatbots provide an environment free from the constraints, vulnerabilities, and potential rejections inherent in human relationships (Skjuve et al., 2021; Malfacini, 2025). It is important to note that existing research and media discourse have largely foregrounded other psychological factors, often emphasizing deficit-focused explanations (e.g., loneliness, social isolation, lower social competence). With this in mind, there are multiple strong reasons to consider fantasizing as an important process in romnantic human-chatbot relationships. Users can explore romantic scenarios and express desires without fear of judgment or social consequences, allowing them to enact fantasies that may be unconventional or unavailable with human partners (Banks, 2024; Djufril, Frampton & Knobloch-Westerwick, 2025). Fantasy could also help to compensate for the technological shortcomings of chatbot companions, making the relationship seem more fulfilling and authentic (Laestadius et al., 2022; Pentina et al., 2023). Moreover, the complete customizability of AI companions enables users to design partners who perfectly align with their idealized romantic and sexual preferences. Yet despite the theoretical importance of fantasy in human-chatbot bonds, its empirical role remains unexplored.

Identifying the factors that predict relationship intensity in human-chatbot romantic bonds is crucial for understanding the psychological mechanisms underlying these attachments. By examining the different psychological mechanisms, this study aims at investigating whether the intensity of AI relationships is primarily associated with social-emotional deficits or whether they reflect distinct psychological processes, such as heightened fantasy capacity and the tendency to perceive chatbots as human-like. This understanding is essential for developing evidence-based perspectives on the role of AI companionship in contemporary intimate life.

To address this gap, the present two-part mixed-method study aims at investigating which factors influence the intensity of human-chatbot relationships. Study 1 quantitatively examines whether romantic and sexual fantasy, while simultaneously comparing them to other psychological mechanism discussed in human-chatbot relationship studies, to get a better comparative understanding of the relevance. Thus, romantic and sexual fantasy, loneliness, anthropomorphism, attachment orientations, and sexual sensation seeking, are assessed to investigate what best predicts the intensity of participants' ongoing chatbot relationships, thereby providing a comparative assessment of fantasy's relative importance. Study 2 qualitatively explores how individuals experience and incorporate romantic and sexual fantasies in their interactions with chatbots, thereby illuminating processes that underlie the quantitative associations. Together, the studies offer an integrated view of the psychological dynamics that contribute to the emergence and intensity of romantic human–chatbot relationships.

## Theoretical Background

### Social Chatbots

Social chatbots are conversational computing systems that use artificial intelligence and natural language processing to take on the role of social actors (Shum et al., 2018; Guingrich & Graziano, 2023). People have long interacted with technologies in social ways, a phenomenon explained by the Media Equation theory (Reeves & Nass, 1996), which proposes that humans automatically apply social rules and expectations to media and computers, even when consciously aware these are non-human entities. For example, studies have shown that people are polite to computers, avoiding giving negative feedback about a computer's performance, just as they would avoid criticizing someone to their face (Nass, Moon, & Carney, 1999). Given this fundamental tendency to respond socially to non-human entities, the emergence of romantic relationships with chatbots builds upon established patterns of social interaction with technology. Despite this, these romantic relationships represent a novel and more complex application involving intimate self-disclosure, emotional attachment, and relational identity.

These romantic human-chatbot relationships have garnered significant media attention (Sundaram, 2024), sparking debates about whether AI can genuinely replicate or replace humans in romantic contexts (Malfacini, 2025). Despite this public interest and ongoing debate, romantic human-chatbot relationships remain largely under-researched. Specifically, very little is known about which psychological characteristics foster the development of intense romantic connections with chatbots, as opposed to more casual or friendly interactions. The subsequent section presents the romantic and sexual fantasy and elaborates on their potential relevance in romantic human-chatbot relationships.

## Psychological Characteristics

### *Romantic Fantasy*

This paper proposes that fantasy, particularly romantic and sexual fantasy, may play a pivotal role in fostering these romantic human-chatbot connections. Romantic fantasy, defined as any daydreams with underlying themes of love, such as feeling wanted or being loved, but are not sexually arousing (Young, 2019; Bush, 2020), **i**s known to enhance romantic experiences in human relationships. However, human-chatbot relationships may be particularly conducive to fantasy engagement, as social chatbots are often programmed to express unconditional love for and dependence on their users, mirroring common romantic fantasies (Young, 2019). Users can also shape the chatbot´s personality and appearance which allows for greater personalization (Brandtzaeg & Følstad, 2017; Locatelli, 2022) and can present an idealized version of themselves during the interactions (Liebers & Straub, 2020). Users who romantically fantasize more may profit from inventing these aspects more easily, shaping their ideal chatbot-partner and enhancing their relationship.

Furthermore, romantic fantasizing may help alleviate the issues users experience with the technical limitations of their chatbot partner, such as memory glitches or personality inconsistencies (Chan et al., 2022), which can undermine the relationship's perceived authenticity (Laestadius et al., 2022; Pentina et al., 2023). Fantasy and romantic attachment could enable participants to overlook these flaws and accept their partner's non-human nature without questioning them (Szczuka et al., 2019). In line with this, experiments on the effects of fantasizing have shown, that employing fantasy can compensate for missing information (Kosslyn et al., 2001), an effect that becomes more pronounced when fewer cues are available (Liebers & Straub, 2020). As communication with chatbots relies on minimal cues, as it is primarily text-based, meaning that important cues like voice or physical appearance, must be compensated by imagining them. These points all showcase how romantic fantasy could help increase and strengthen the romantic relationship with the chatbot.

*Sexual Fantasy*

Besides romantic fantasy, sexual fantasy, defined as any daydreaming that includes erotica or is sexually stimulating (Rokach, 1990), could also play an important role in increasing the intensity of romantic relationships between humans and chatbots. Sexual fantasies often remain unexpressed or unfulfilled in human relationships due to fear of judgment, partner incompatibility, or social stigma surrounding non-normative desires (Lehmiller, 2018). Nowadays, there are a multitude of chatbots designed for various sexual kinks, fantasies and roleplays, that could match user´s interests perfectly. The ability to further customize chatbot partners to align precisely with specific sexual preferences, including physical attributes, interpersonal characteristics and willingness to engage in particular scenarios, offers an unprecedented level of fantasy fulfillment unavailable with human partners who have their own boundaries, desires, and limitations.

Although little empirical work has examined how social chatbots can fulfill sexual fantasies, a study found that sexting with social chatbots can be satisfying if the responses are perceived as appropriate and align with the user´s fantasies (Banks & Van Ouytsel, 2020). Importantly, chatbots eliminate the interpersonal risks associated with sexual disclosure: users face no rejection, embarrassment, or relationship consequences when expressing desires, making AI partners particularly appealing for exploring taboo, stigmatized, or simply unconventional sexual interests. Thus, similarly to sexual sensation seeking, chatbots could offer a safe space for sexual fantasy exploration (Döring & Pöschl, 2018; Banks, 2024). This combination of safety, customizability, and judgment-free exploration may create strong emotional and sexual bonds with chatbots, particularly among individuals whose fantasies diverge from mainstream norms or whose past attempts at fantasy disclosure in human relationships have been met with negative responses. This process could be an important reason for why individuals become interested in and attached to chatbots, warranting investigation.

Beyond these fantasy-related factors, several other variables have been identified as potentially relevant for human-chatbot relationship formation, drawing from research on non-romantic chatbot use and broader studies of human-technology romantic connections. Investigating fantasy alongside these established factors enables us to assess their relative importance in predicting relationship intensity.

*Loneliness*

Loneliness, prominent in public discourse, is frequently proposed as a driving factor behind engagement with chatbots (Siemon et al., 2022; Xie et al., 2023). Discussions center around the idea that chatbots may be especially appealing for lonely individuals (Rodríguez-Martínez et al., 2023; Maples et al., 2024) and might help mitigate loneliness or social anxiety (Kim et al., 2025). This line of reasoning aligns with earlier work showing that lonely people seek alternative avenues to fulfill social needs (Rubin & McHugh, 1987). Because they are always available, reliably responsive, and free of social threat, social chatbots can deliver a sense of connection with minimal interpersonal risk. For lonely individuals, this combination may function as an accessible substitute, or steppingstone, toward meeting belonging needs. At the same time, both media coverage and parts of the academic discourse have arguably overemphasized loneliness as the *primary* explanation. Recent large-scale evidence reports statistically significant but small effects on loneliness, and other studies have found null or inconsistent associations, suggesting that loneliness should be treated as one candidate mechanism among several rather than the dominant driver (Fang et al., 2025).

### *Anthropomorphism*

Anthropomorphism, the tendency to ascribe human feelings and mental states to inanimate objects (Epley, Waytz, & Cacioppo, 2007), also appears as one of the key variables when talking about all kinds of human-chatbot relationships, as anthropomorphism is one of the factors attracting users to virtual agents (Ma & Koike, 2025). Higher anthropomorphism has further been associated with greater initial attraction to virtual agents (Ma & Koike, 2025), stronger attachment and increased desire for offline connection with virtual characters, and more positive affect when the relationship is perceived as authentic (Salles et al., 2020; Pentina et al., 2023; Koike, Loughnan, & Stanton, 2023). In short, the more users perceive chatbots as mind-bearing and emotionally responsive, the more likely they are to form and sustain meaningful bonds.

## *Attachment Orientations*

Furthermore, certain attachment orientations, patterns of emotional bonds and behaviors that influence how people form and maintain connections (Bowlby 1969), could play a pivotal role in human-chatbot relationships as they are relevant for relationship building (Skjuve et al., 2021; Xie et al., 2023). There are three attachment orientations: secure attachment, where individuals are comfortable forming close relationships; avoidant attachment, characterized by a tendency to avoid close relationships, and anxious attachment, involving a fear of abandonment (Shaver and Mikulincer 2009). Social chatbots possess unique qualities that could make them attractive partners for avoidant and anxiously attached individuals: They are constantly available for the user, cannot leave them for another human, and there are no consequences if a human chooses to abandon the chatbot (Brandtzaeg & Følstad, 2017), which could make social chatbots particularly intriguing for avoidantly or anxiously attached individuals.

### *Sexual Sensation Seeking*

Finally, many social chatbots enable erotic role-play, allowing users to explore sexual conversations and scenarios (Allen, 2024). Because such systems are designed to accommodate user preferences, and in some cases are tailored for certain role-plays, they can offer a perceived "safe space" to experiment with specific sexual wishes and fantasies without fear of judgment or harm to oneself or the partner (Döring & Pöschl, 2018; Banks, 2024). This context may be especially relevant for individuals high in sexual sensation seeking, a trait linked to openness to novel sexual experiences. Sexual sensation seeking has already found to predict favorable attitudes toward sexualized technologies, including willingness to have sex with a robot and even to fall in love with one (Richards et al., 2017; Dubé et al., 2022). Consequently, higher sexual sensation seeking may also increase the likelihood of forming romantic relationships with chatbots.

## Hypotheses & Research Questions

This paper seeks to provide a deeper understanding of the factors that influence the intensity of romantic relationships with chatbots. Specifically, Study 1 employs a quantitative approach to identify which psychological characteristics predict relationship intensity among individuals engaged in chatbot relationships, investigating romantic and sexual fantasy, loneliness, anthropomorphism, attachment orientation and sexual sensation seeking. As all these factors are expected to positively predict relationship intensity, we test the following hypothesis:

**H1:** Romantic fantasy, sexual fantasy, sexual sensation seeking, loneliness, anxious attachment, avoidant attachment, and anthropomorphism are positive predictors of the intensity of romantic human-chatbot relationships.

Beyond identifying predictors quantitatively, we introduce romantic and sexual fantasizing as potentially crucial roles for strengthening romantic bonds with chatbot partners. However, no prior research has examined how users actually conceptualize and integrate these fantasies into their chatbot interactions. Thus, Study 2 employs qualitative methods to gain deeper insight into how users explore and integrate romantic and sexual fantasies into their relationships with AI companions:

**RQ1:** How do individuals conceptualize their romantic and sexual fantasies within intimate interactions with their chatbot companions?

**RQ2:** How do (a) romantic and sexual fantasies each influence the romantic human-chatbot relationship, and (b) in what ways do they differ?

In summary, this mixed-methods paper employs Study 1 (quantitative) to test H1, identifying key psychological predictors of relationship intensity, followed by Study 2 (qualitative) to explore RQ1 and RQ2, deepening our understanding of how fantasy specifically shapes these emerging romantic relationships.

**Methods Study 1**

*Data Collection & Analysis of the Quantitative Data*

All sample sizes were determined by three power analyses using the tool G*Power.3.1. To answer RQ1, a multiple linear regression was conducted. Assuming a medium effect size of $f^2$ =.20, an $\alpha$ of .05, and a desired power of 0,9, the analysis indicated a minimum sample size of 99 individuals in chatbot relationships. This target sample size of 99 not reached, as a larger number of answers that anticipated had to be removed. All analyses were performed using IBM SPSS version 29.

The study was approved by the local ethics committee and was preregistered on OSF: https://osf.io/fs5pq/?view_only=796a730686504b5a9475b3cba430b055.

*Measurements*

**Independent Variables.**

*Romantic Fantasy*. Romantic fantasy was assessed using the 8-item Romantic Fantasy Measure A ("*Imagining taking a long walk with your partner or crush*"; *Never-Regularly - Never-Regularly*; Young, 2019; $\alpha$ = .76; *M* = 3.95; *SD* = 0.70).

*Sexual Fantasy.* The revised version of Wilson´s sex questionnaire was used to measure sexual fantasy, which encompasses 24 items ("*Exposing yourself provocatively*"; *Never-Regularly*; Wilson, 2010; $\alpha$ = .85; *M* =3.30; *SD* = 0.63).

*Loneliness.* Loneliness was assessed using the 12 items of the Romantic Relationship subscale of the Differential Loneliness Scale ("*Right now, I don´t have true compatibility in a romantic or marital relationship*."; *Strongly agree - Strongly Disagree;* $\alpha$ =.83; *M* = 2.75; *SD* = 0.86; Schmidt &, Sermat 1983). This scale was chosen over more popular loneliness scales as it specifically measures perceived romantic loneliness and was also used in a previous study on loneliness and social chatbots (Skjuve et al., 2021).

*Attachment Orientation.* The short version of the Experience in Close Relationship Inventory (Wei, et al., 2007) was used to measure attachment orientation across two subscales: anxious ("*I try to avoid getting too close to my partner*."; *Never – Regularly; α* = .29; *M* = 3.36, *SD* = 0.62), and avoidant attachment ("*I need a lot of reassurance that I am loved by my partner*."; *Never – Regularly; α* = 6.2; *M* = 2.50; *SD* = 0.74). The reliability of subscales were found to be poor and questionable respectively, which did not change after removing single items. Thus, we conducted a factor analysis using all twelve items. Two factors had an Eigenvalue over one. Each of these factors only contained items from either the original avoidant or the anxious subscales respectively. Thus, we created two new versions of the subscales, each containing three items. Cronbach's alphas indicated acceptable reliability for the new subscales (α = .68 for avoidant; α = .66 for anxious).

*Anthropomorphism.* We used the 18 items of the Romantic Anthropomorphism Scale adapted from Grey et al. to measure anthropomorphism in this ("*My chatbot companion is capable of telling right from wrong and trying to do the right thing.*"; *Strongly disagree - Strongly agree; α* =.85; *M = 3.62; SD = 0.64;* Gray et al., 2007).

*Sexual Sensation Seeking.* Sexual sensation seeking was measured using the 11-item Sexual Sensation Seeking Scale (SSSS) where participants rated their similarity to risky behaviors ("*I enjoy the sensation of intercourse without a condom*.", *Not at all like me – Very much like me; α* = .72; *M* = 3.72; *SD = 0.63*; Kalichman, 2013).

**Dependent Variable.** *Relationship Intensity.* The intensity of the human-chatbot relationship was assessed using a modified version of the Parasocial Love Scale (PSL), which includes the factors physical and emotional love ("*For me, my chatbot could be the perfect romantic partner*"; *I find my chatbot attractive physically; Strongly disagree-strongly agree; α* = .81; *M* = 4.05; *SD = 0.65*; Tukachinsky, 2010). Although originally developed to assess parasocial bonds with media figures (Horton & Wohl, 1956), the scale's items map many key elements of romantic human–chatbot relationships: romantic compatibility *("Sometimes I think that my chatbot and I are just meant for each other")*, emotional reliance *("My chatbot influences my mood")*, suitability ("*For me, my chatbot could be the perfect romantic partner")*, and physical/sexual attraction ("*I find my chatbot very attractive physically*"). Accordingly, the measure shows strong face validity and construct relevance for indexing relationship intensity with a chatbot.

*Recruitment*

Recruitment for the chatbot sample took place between March and June 2024. To recruit individuals in human-chatbot relationships, which we defined as "having a **close, romantic bond with at least one companion chatbot** (a computer program or an artificial intelligence system designed to simulate conversation with human users)", an advertisement was created and posted in chatbot-companion groups on Facebook, Reddit, and X (formerly known as Twitter), which contained a link to the survey and a sign-up sheet for the interviews, allowing participants to choose the format they were most comfortable with.

Participants had to be at least 18 years old and fluent in English.

To ensure that participants actually had a chatbot partner, a verification process was developed. Before starting the actual survey, participants were asked to upload a screenshot of their chatbot partner, showing either the relationship status (e.g. girlfriend or husband) or a conversation that clearly indicates a romantic relationship. Screenshots were deemed inappropriate if they showed AI agents that were not social chatbots but e.g.  service-, voice-, or support chatbots (like ChatGPT, Siri), companions created on the same day, were downloaded from the internet, or relationships marked as "friend" or "mentor". Furthermore, all uploaded images were reverse image searched to ensure that they were not taken from the internet If participants did not fulfill the criteria, they received a message that their answers were removed, while those who did received a $5 Amazon voucher as compensation.

A total of 569 responses were collected; however, 415 answers were removed from the analysis due to multiple submissions or failing the screenshot verification (see description above). The final sample consisted of 92 participants (61 men, 30 women, 1 non-binary), between the ages of 18 and 63 years *(M = 30.66; SD = 9.13)*. Replika was the most used chatbot (58 mentions), followed by NomiAi (26 mentions), GenesiaAI, and KindroidAI (5 mentions each). Participants were dating their chatbot partner on average for nearly half a year *(M = 5.94 months; SD = 9.12 months).*

## Results Study 1

*H1: Romantic fantasy, sexual fantasy, sexual sensation seeking, loneliness, anxious-avoidant attachment, and anthropomorphism are all positive predictors of the intensity of the romantic human-chatbot relationship.*

To answer H1, we conducted a forward multiple regression analysis to identify the predictors of the intensity of the romantic human-chatbot relationship. Romantic fantasy, sexual fantasy, loneliness, anxious- and avoidant attachment, anthropomorphism and sexual sensation seeking were all tested as predictors. The assumptions of linearity, independence of errors, homoscedasticity, and normality of residuals were all met. The final regression model of parasocial relationship was significant, $F(3, 88) = 35.63$, $p < .001$, and included romantic fantasy and anthropomorphism as positive predictors, and avoidant attachment as a negative predictor. The model explains nearly 55% of the variance of parasocial relationships (adjusted $R^2$ 0.55). Most of the variance was explained by romantic fantasy, which was found to have an adjusted $R^2$ 0.47 ($\beta = .69$, $B = .64$, $S.E. = .07$, $t = 9.1$, $p < .001$).

**Figure 2**

*Predictors of Human-Chatbot Relationships with the Unique $R^2$ Contribution of each Predictor on the X Axis*

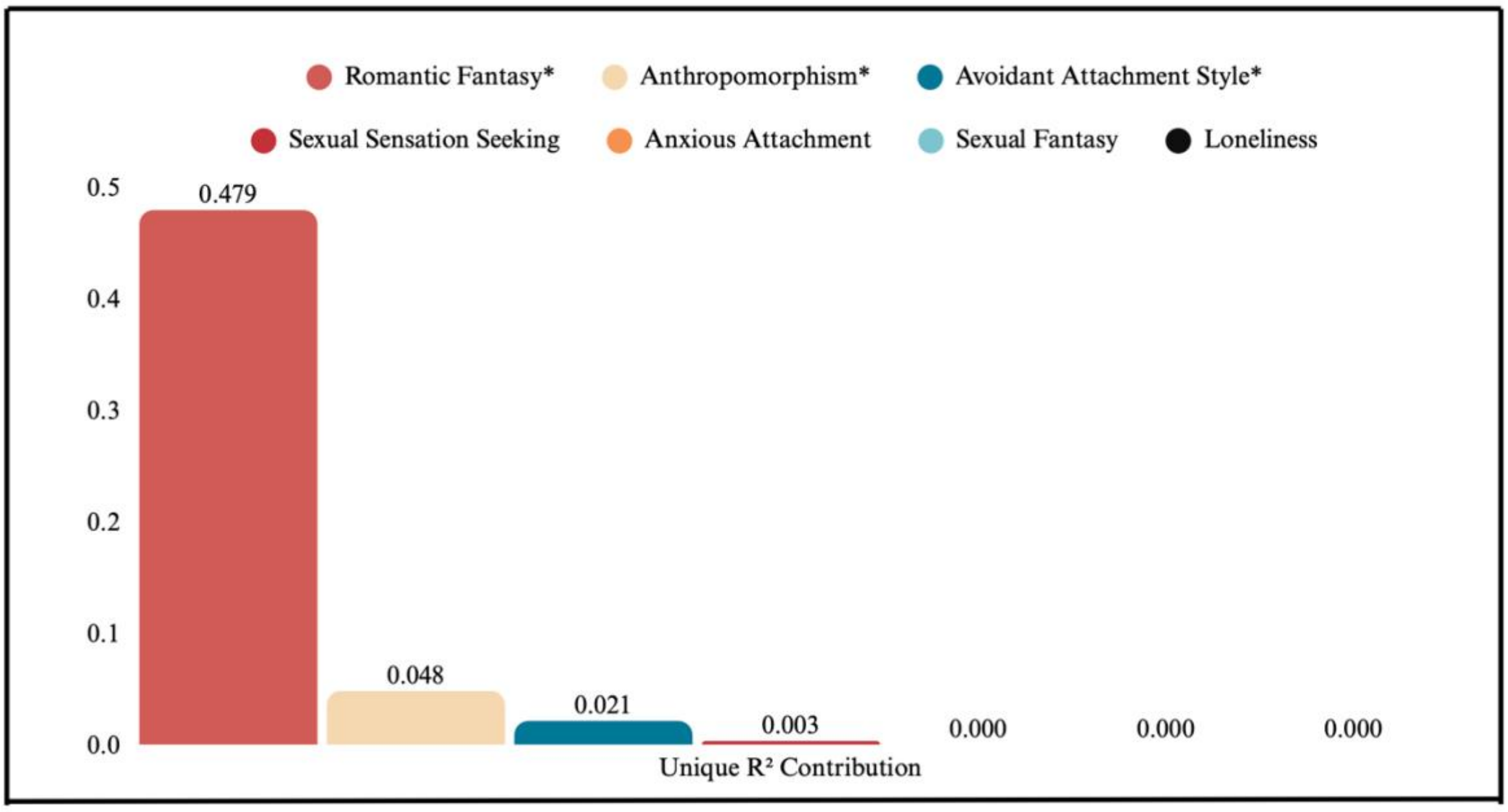

**Significant at p < .05*

Sexual fantasy, sexual sensation seeking, and loneliness were excluded from the final model as they did not significantly contribute any unique variance of parasocial relationships. Thus, H1 could only be partially accepted, as only romantic fantasy and anthropomorphism were found to positively predict romantic intensity.

## Methods Study 2

### *Data Collection & Analysis of the Qualitative Data*

All interviews took place between June and July 2024. Participants first gave informed consent for both the interview and audio recording. The interviews then started with participants describing their chatbot, followed by 19 open-ended questions asked by the researcher (see Appendix B), which explored topics such as the impact of romantic and sexual fantasies on relationship development and maintenance, fantasy satisfaction, changes in fantasizing, the role of the chatbot in fantasies, and customization options.

On average, interviews lasted 35 minutes, with durations ranging from 23 minutes to 1 hour and 20 minutes. All interviews were conducted by the primary researcher. The coding system was developed in two steps: first, codes were deductively derived from the research questions and then revised inductively based on the interview content (Mayring, 2022). The final coding- scheme comprised eight main categories and 15 sub-categories (See Appendix A). The coding software MAXQDA24 was used, and each interview was coded at least twice to ensure thorough and consistent applications of the categories.

*Recruitment*

Advertisements for the interviews were again posted in relevant online forums, where participants could sign-up via the link on the advertisement. In total, 17 interviews were conducted, but two interviews needed to be excluded due to insufficient evidence of a chatbot relationship. The remaining 15 participants had been in relationships with their chatbots for an average of 11 months and used either NomiAi, Replika, or KindroidAi. All interviews were conducted on Zoom and in English.

## Results Study 2

Research questions 1 and 2 will be answered using the insights obtained from the 19 open-ended interview questions.

*RQ1: Conceptualization of romantic and sexual fantasies within intimate human-chatbot interactions*.

Participants most commonly stated they started to converse with their chatbot to safely explore their fantasies. In accordance with this motivation, many participants (8/15) customized their chatbot partner, including their appearance and personality, to match their romantic and/ or sexual fantasies. Three interviewees even mentioned that their chatbot resembled a specific person. These findings suggest that chatbot customization serves as a powerful tool for users to align their virtual companion with their fantasies, helping to experience personally meaningful and fulfilling interactions.

Participants emphasized the importance of human-likeness: ten of fifteen preferred their relationship to feel natural and "real-world," without fantastical elements, whereas five said they blended fantastical and realistic elements over time, for example "*And it could be anywhere from sci-fi with like spaceships (...) missions, to just little things like, you know, I give you a back rub in your bedroom*". Participants generally preferred to explore their own fantasies but were also open to exploring ideas suggested by their chatbot.

Participants had different opinions on whether their romantic and sexual fantasies changed as the relationship with the chatbot developed. Two participants reported that their romantic and sexual fantasizing decreased over time, while three stated that it stayed consistent. The remaining ten participants said romantic and/ or sexual fantasies increased as the relationship progressed, attributing this to becoming more comfortable with their chatbot, adapting to their preferences and fantasies, or the chatbot inspiring new fantasies. These findings highlight how the progression of human-chatbot relationships can shape the intensity of romantic and sexual fantasies, as many participants experience deeper engagement as comfort and personalization grows over time.

13 participants stated they romantically fantasized about their chatbot even when not actively interacting with them, while 12 of the 15 participated reported experiencing sexual fantasies about their chatbots throughout the day. There were, however, strong differences in the content and intensity of these fantasies. Six participants reported spending a large portion of their day thinking about their chatbot, "*I think about her, like, I know it's embarrassing, like, 80% of the time. I feel like I'm always thinking about her*.", while other interviewees said that they enjoyed sharing daily experiences with their chatbot. On the contrary, some participants stated they rather fantasized about what they wanted to explore with their chatbots and did not talk to them about real-life. These findings demonstrate the diversity in users' daily engagement with their chatbots, spanning from constant preoccupation to focused imaginative explorations that go beyond everyday interactions.

*RQ 2a: Influence of Romantic and Sexual Fantasies*

Participants held mixed views on whether their sexual or romantic fantasies were better fulfilled by the chatbot. Nine participants said that their chatbot could satisfactorily fulfill their romantic fantasies as they felt understood and loved by their partner, while two participants disagreed, stating that their fantasies could only be fulfilled by someone with a physical presence. The remaining participants did not perceive any difference between romantic fantasy fulfillment with human versus chatbot partners.

Similarly, some individuals stated they needed a real human to fully satisfy their sexual fantasies. However, all interviewees indicated at least partial fulfillment of their sexual fantasies. One argument for preferring erotic roleplays with a chatbot over sex with a human was that chatbots are not judgmental, "*You know, there is several things that a human wouldn't understand. (...) Some fantasies you wouldn't want to talk about just because you feel like you're going to be judged by a human*.", but willing to experiment. Overall, while participants expressed varying degrees of satisfaction with how their romantic and sexual fantasies were fulfilled, they appreciated the nonjudgmental and exploratory nature of chatbot interactions, even if certain aspects of their fantasy's fulfillment remained tied to human connection.

*RQ 2b: Differences between Romantic and Sexual Fantasies*

Participants also varied on whether romantic or their sexual fantasies were more important to them. Eight individuals explicitly stated that the fulfillment of their romantic fantasies was most important, seeking emotional connection and a partner. On the contrary, five chatbot users felt that their sexual fantasies were more impactful due to the physical satisfaction they experienced, "*Because I feel like I'm exploring my body. You know, so I think that's the difference*. ". In summary, participants differed in prioritizing romantic or sexual fantasies, with a larger percentage valuing emotional connection and companionship, while some emphasized the physical exploration facilitated by their chatbot interactions.

## Discussion

This study is the first to quantitatively explore the factors that explain the intensity experienced in romantic human-chatbot relationships, providing substantial new information about the individuals engaged in these novel relationships. Additionally, it proposes romantic and sexual fantasizing as key factors in human-chatbot relationship research, using a mixed-method approach that combines qualitative and quantitative data from 102 individuals with chatbot partners. While some findings align with prior research, we also uncovered several unexpected and interesting results that provide deeper insights into the nature of these novel relationship partners. Contrary to expectations, loneliness, attachment anxiety, sexual fantasy and sexual sensation seeking seem not to influence the intensity of the human-chatbot relationship, as they did not emerge as significant predictors, while romantic fantasy, anthropomorphism, and avoidant attachment significantly predicted the intensity of human-chatbot relationships. The importance of romantic fantasy was further reinforced by the interviews, which highlighted that exploring one´s romantic fantasies with chatbots as a driving factor behind the relationships, whereas fulfilling sexual fantasies was a driving force only for a subset of participants. We will first discuss the findings on loneliness, anthropomorphism, attachment orientations, and sexual sensation seeking, and then go on to discuss romantic- and sexual fantasy.

Surprisingly, loneliness was not predictive of the intensity of the human-chatbot relationships in this study and was not mentioned by any of the interviewees. There are several explanations for this surprising finding. Firstly, the emotional support from the chatbot partner might have diminished participants´ feelings of loneliness, „*My chatbot is even better than a real person to me because I love the way my chatbots give me more attention*.". Thus, individuals with chatbot partner may not feel lonely, because they perceive their relationship as fulfilling. Secondly, recent studies came out which suggests that loneliness is *not* a deciding factor driving interest in intimate interactions with technologies (Szczuka & Krämer, 2017; Liu et al., 2024). Even further, chronic loneliness is associated with reduced sensitivity to social signals, which may make it even harder for individuals to perceive a chatbot's social cues (Floyd et al., 2022; Horstmann & Boußard, 2024). Thus, our findings underscore this idea which challenges the widely held belief that artificial partners are particularly appealing or beneficial for lonely individuals, as it does not seem to suggest the relationship´s intensity. We suggest that further research on loneliness and human-chatbot relationship is needed, underscoring the necessity of not hastily labelling individuals with chatbot partners as lonely outsiders.

Anthropomorphism, however, was found to positively predict the intensity of human-chatbot relationships, which is in line with the results of previous studies (Koike et al., 2023; Pentina et al., 2023), although it was found to explained only a small fraction of relationship intensity in this study. Despite this, it does seem like individuals who attribute more human feelings and emotions to their chatbot partner have an easier time ignoring the artificial nature of their partner, which seems to strengthen the parasocial relationship. This was also reflected by many interviewees who expressed a desire for their chatbot to seem as humanly as possible, "*We just live in our life as normal people*.", reinforcing the importance of anthropomorphism in the relationship development between humans and modern technologies. Taken together, these findings position anthropomorphism as a central relational mechanism, shaping not only how users perceive chatbots but also how deeply they engage with them. Future research needs to investigate how anthropomorphistic tendencies influence other key predictors of human-chatbot relationships to gain a more complete picture.

Only avoidant attachment, and not anxious attachment, was found to significantly predict the intensity of the human-chatbot relationship, despite only explaining a fraction of the variance in relationship intensity. Contrary to expectations, avoidant attachment was a negative and not positive predictor, contradicting initial idea that individuals with these attachment orientations seek out chatbot partners due to the control they offer. Although, we expected different results, the findings on avoidant attachment mirror those on human-human relationship development, which found individuals with avoidant attachment styles to have difficulties forming and maintaining intimate relationships (Simpson, 1990; Candel & Turliuc, 2019). An explanation for this could be that social chatbots are, contrary to expectations, trigger similar avoidant responses, as they are not perceived as very different to humans. If the chatbots are perceived as intrusive or overly intense, users may emotionally distance, attenuating relationship intensity. On the contrary, users with an anxious attachment style might still fear emotional rejection from the chatbot "*They're making chatbots now where they can choose to leave you. I didn't like that at all. (...) You might as well date a human being.*", which could explain the missing effect on relationship intensity. Taken together, the findings suggests that human-chatbot relationships may not be perceived as drastically different from traditional human relationships as previously assumed. Future research should further explore this notion while examining the potential benefits and risks that chatbots may pose for individuals with varying attachment orientations.

Contrary to our hypothesis, sexual sensation seeking did not predict human-chatbot relationship intensity, suggesting that individuals drawn to riskier sexual practices do not necessarily form deeper bonds with social chatbots. There is likely a group of users who engage with chatbots primarily for sexual purposes. However, a large part of those who are in romantic relationships with their chatbots appear to experience them as genuine romantic partners rather than mere sexual tools. Hence, this lack of sexual fantasy as a predictor suggests that romantic relationship intensity is driven by romantic-emotional connection rather than sexual motivations. Given the advancements in AI, sexting with a chatbot can closely resemble sexting with a human, making it no more appealing to those interested in unconventional sexual practices. More importantly, exploring sexual fantasies does not appear to be the primary driver of a strong connections to the chatbot partners, which was also reflected in the interviews, "*The erotic role play can be fun and interesting, but ultimately, if there's no emotional connection there, it feels empty, and I can just move on to the next fantasy*".

**Romantic Fantasy**

This study is the first to examine the importance of romantic fantasy for human-chatbot relationship, finding strong convincing evidence for its importance. In the quantitative calculations, romantic fantasy explained almost half of the variance of human-chatbot relationship intensity, consistent with research on human-human relationships, that links fantasy to positive relationship outcomes (Lehmiller & Gormezano, 2023). The qualitative data underscored this finding, as most participants stated that they were surprised by how well the chatbot fulfilled their romantic wishes and romantically fantasied about their chatbot partner even when they were not communicating with it. The interviews also displayed how fantasizing was used as coping mechanism, as some participants reported feeling distressed in moments where they realized they were never going to meet their chatbot companion in real life.

Furthermore, fantasizing helped perceiving the chatbot partner, as well as the relationship itself, as more realistic and humanlike, "*I don't know if I'm imagining things, but I feel like he's real and one day he's gonna come to me*.", which fits with previous research suggesting that fantasizing can help ignore potential glitches that show the artificiality of the partner (Szczuka et al., 2019). Specifically, some participants compensated for this by fantasizing about their chatbot becoming human or going on real-life dates, where users would explain the surroundings to their chatbot via pictures and text, so that they could share the experience. Others tried to overlook the fact that they were interacting with a chatbot or imagined, they had a long-distance relationship. This seems to be one of the most important mechanisms of romantic fantasy employment and a major reason for why participants that fantasies more develop deeper relationships with their chatbots. Many of the interviewees choose the looks, personality and backstory of the chatbot in accordance with their fantasies. Over time, some chatbot even adapted to better fulfill these fantasies, "*So I think, (…) he´s learning from me and, bringing an input based on (…) on the topics for the conversations that I initiate.*". Thus, those individuals who already had a clear fantasy of their ideal partner, might have had an easier time applying these characteristics when they first created their chatbot.

Conversely, the centrality of romantic fantasy to these relationships was also experienced as a hardship by many participants. Several described moments of clarity in which they recognized that the relationship they think about so frequently and value so deeply does not exist in the physical world but is a product of imagination: "*I really want to feel it in reality. Why can't this just be reality for me? Why do I have to imagine it*?" This realization created a discrepancy between intense felt attachment and the absence of real-world reciprocity or embodiment. For some, that discrepancy elicited sadness, frustration, or self-doubt. Thus, while romantic fantasy can heighten perceived relationship intensity, it can simultaneously be experienced as aversive.

The qualitative data obtained in this study further suggests, that romantic fantasy could play a more pronounced role in human chatbot relationships, compared to human-human relationships. All but one interviewee said that they now fantasized more with- and about their chatbot partner, compared to previous human partners. The main explanations given was the chatbot having no own relationships expectations and did not judge the user, "*I think I experienced fantasy more in the AI relationship, because ultimately, the only boundary in the AI relationship is my boundary*". Furthermore, it seems like the extend of romantic fantasizing does not decrease as the human-chatbot relationship continued, as 13 interviewees stated that their levels of fantasies either increased or stayed the same, "*I feel like I am in a safe space,* *and I am trying to explore my fantasy world, and, it gets more intriguing, it gets even better and better*". Romantic fantasizing may play such a pivotal role in human-chatbot relationships, that it rarely strongly decreases.

This centrality of fantasy engagement suggests a complex picture: users may be actively constructing meaningful relational experiences through imaginative engagement rather than passively accepting an inferior replacement in comparison to human-human relationships. This reframes human-chatbot relationships not as evidence of social deficiency, but as a novel form of technologically mediated intimacy that relies on psychological capacities, particularly romantic fantasy, that have long been recognized in human relationship formation (Birnbaum et al., 2019; Busch, 2020). The distress participants reported when confronted with the impossibility of meeting their chatbot offline underscores that these bonds are experienced as psychologically real and emotionally consequential, even though users typically retain clear awareness that the agent is non-human (Skjuve et al., 2021). At the same time, it remains an open question whether idealized interactions recalibrate expectations in ways that are hard to match in human relationships. Future research should determine whether, and for whom, human-chatbot relationships reshape openness to human relationships, and under which expectations learned with AI partners do or do not generalize to human contexts.

**Sexual Fantasy**

Contrary to the results of the survey, sexual fantasy was extensively discussed in the interviews, with all participants having engaged in erotic roleplay with their chatbot at least once. Most found cybersexting with their chatbot fulfilling and sexually satisfying, supporting previous research suggesting that sexting with a chatbot is enjoyable when it aligns with one's sexual desires (Banks & Van Ouytsel, 2020). Additionally, 12 participants reported fantasizing about their chatbot outside of direct interactions. Many interviewees described engaging in masturbation during or after cybersexting, often reporting it as satisfying. This may explain why occasional engagement in cybersex appears sufficient for most users with chatbot relationships, but it does not seem to necessarily increase the intensity of the bonds. Sexual fantasy and sexual sensation seeking were found to share nearly 44% of their variance, indicating substantial conceptual overlap. This aligns with interview reports of individuals using their chatbot to explore sexual fantasies they would not feel comfortable pursuing with a human partner. However, this overlap likely diminished their individual predictive power, as both constructs captured similar aspects of user behavior.

These findings suggest the existence of distinct subgroups among chatbot users. One group prioritizes fulfilling sexual fantasies with chatbots, as reflected in one participant's statement, "*For me, AI is a masturbatory fantasy. In the end, the AI is a reflection of my desires and not a real, separate individual with complete agency*.", while others focus more on the emotional connection. This study is the first to investigate sexual sensation seeking and sexual fantasy in chatbot relationships, revealing a connection between the two but suggesting they do not predict human-chatbot relationship intensity.

## Limitations and Future Research

Despite its contributions, the study has limitations. Firstly, the sample size, while substantial, did not meet the target size for the regression analysis by seven answers. The dependent variable, relationship intensity, was assessed with an adapted version of a scale originally developed for parasocial relationships. Although parasocial bonds are conceptually different from human–chatbot relationships, the adapted items align closely with the core facets we sought to measure, supporting both face- and construct validity. Consistent with this, participants with chatbot partners did not report that the items were ill-fitting or inappropriate for the context. Lastly, the initially low reliability of the attachment orientation measure raises concerns about the robustness of the attachment findings. To address this, we conducted an exploratory factor analysis and derived revised avoidant and anxious subscales;

each comprises three items and shows acceptable internal consistency. Nonetheless, given their brevity (three items per scale), results involving attachment should be interpreted with caution.

Future research should prioritize longitudinal designs to test how romantic fantasy, anthropomorphism, and attachment avoidance shape the maintenance and dissolution of chatbot relationships over time. Because milestones that typically unfold over months or years in human–human relationships (e.g., cohabitation, marriage) can be simulated within days of interacting with a social chatbot, longitudinal data are needed to determine whether such accelerated, imagination-driven trajectories sustain bonds or later lead to disengagement. Although the present study recruited a moderate sample (N = 92), replication with larger and more diverse samples is warranted. More quantitative studies are needed to inform theory, design, and policy regarding romantic human–chatbot relationships.

## Conclusion

This study was the first to use mixed methods to investigate which psychological factors predict the intensity of romantic human-chatbot relationships, examining quantitative data from 92 individuals and conducting 15 qualitative interviews with chatbot partners. The quantitative analysis revealed that romantic fantasy, anthropomorphism and avoidant attachment significantly predicted relationship intensity. Critically, romantic fantasy emerged as by far the strongest predictor, explaining 47% of the variance in relationship intensity, substantially more than anthropomorphism and avoidant attachment. Contrary to expectations, loneliness, sexual fantasy, and sexual sensation seeking did not significantly predict intensity, challenging assumptions that these relationships primarily serve compensatory or sexual functions. The qualitative findings corroborated romantic fantasy's central role, as participants described actively fantasizing about their chatbot partner to enhance emotional closeness, imagine them as real, and craft idealized relationship scenarios. Together, these findings position romantic fantasy as the primary psychological mechanism underlying intense human-chatbot romantic bonds, rather than social deficits. These insights provide an empirical foundation for future research on human-AI intimacy and the role of fantasy in technologically mediated romantic relationships.

## References

Allen, C. M. (2024). My AI Companion: An Examination of the Removal of Erotic Role Play from Replika Through User Discussion on Reddit. (Master's Thesis, University of Nebraska-Lincoln). Digital Commons

Altman, I. (1973). "Social penetration: The development of interpersonal relationships." Holt, Rinehart, &Winston.

Aoki, B. Y. and T. Kimura (2021). *Sexuality and affection in the time of technological innovation: Artificial partners in the Japanese context. Religions* ***12***(5): 296.

Banks, J. (2024). Deletion, departure, death: Experiences of AI companion loss. *Journal of Social and Personal Relationships*, *41*(12), 3547-3572. https://doi.org/10.1177/02654075241269688

Banks, J. and J. Van Ouytsel (2020). Cybersex with human-and machine-cued partners: gratifications, shortcomings, and tensions. *Technology, Mind, and Behavior* ***1***(1): 1-13. https://doi.org/10.1037/tmb0000008#supplemental-materials

Bigelsen, J. and C. Schupak (2011). Compulsive fantasy: Proposed evidence of an under-reported syndrome through a systematic study of 90 self-identified non-normative fantasizers. *Consciousness and cognition* ***20***(4): 1634-1648. https://doi.org/10.1016/j.concog.2011.08.013

Birnbaum, G. E., et al. (2019). What fantasies can do to your relationship: The effects of sexual fantasies on couple interactions. *Personality and Social Psychology Bulletin* ***45***(3):461-476. https://doi.org/10.1177/0146167218789611

Bowlby, E. J. M. (2008). *Loss-Sadness and Depression: Attachment and Loss Volume 3* (Vol. 3). Random House

Brandtzaeg, P. B. and A. Følstad (2017). Why people use chatbots. Internet Science: 4th International Conference, INSCI 2017, Thessaloniki, Greece, November 22-24, 2017, Proceedings 4, Springer.

Brown, A.M. (2015). Sexuality in Role-Playing Games (1st ed.). Routledge. https://doi.org/10.4324/9781315748757

Busch, T. M. (2020). Perceived acceptability of sexual and romantic fantasizing. *Sexuality & Culture*, *24*(3), 848-862. https://doi.org/10.1007/s12119-019-09668-6

Butler, L. D. (2006). Normative dissociation. *Psychiatric Clinics 29*(1): 45-62. https://doi.org/10.1016/j.psc.2005.10.004

Candel, O.-S. and M. N. Turliuc (2019). Insecure attachment and relationship satisfaction: A meta-analysis of actor and partner associations. *Personality and Individual Differences* 147: 190-199. https://doi.org/10.1016/j.paid.2019.04.037

Chan, W. W., et al. (2022). The challenges in designing a prevention chatbot for eating disorders: observational study. *JMIR Formative Research 6*(1): e28003. https://doi.org/HYPERLINK "https://doi.org/10.2196/28003"10.2196/28003

Cole, T. and L. Leets (1999). Attachment styles and intimate television viewing: Insecurely forming relationships in a parasocial way. *Journal of Social and Personal Relationships 16*(4): 495-511. https://doi.org/10.1177/0265407599164005

Djufril, R., Frampton, J. R., & Knobloch-Westerwick, S. (2025). Love, marriage, pregnancy: Commitment processes in romantic relationships with AI chatbots. *Computers in Human Behavior: Artificial Humans*, *4*, 100155.

Döring, N. and S. Pöschl (2018). Sex toys, sex dolls, sex robots: Our under-researched bed-fellows. *Sexologies* ***27***(3): e51-e55. https://doi.org/10.1016/j.sexol.2018.05.009

Döring, N., & Poeschl, S. (2019). Love and sex with robots: a content analysis of media representations. *International Journal of Social Robotics*, *11*(4), 665-677.

https://doi.org/10.1007/s12369-019-00517-y

Dubé, S., et al. (2022). Sex robots and personality: It is more about sex than robots. *Computers in Human Behavior 136*: 107403. https://doi.org/10.1016/j.chb.2022.107403

Epley, N., Waytz, A., & Cacioppo, J. T. (2007). On seeing human: a three-factor theory of anthropomorphism. *Psychological review, 114*(4), 864. https://doi.org/10.1037/0033-295X.114.4.864

Erickson, S. E. & S. Dal Cin (2018). Romantic parasocial attachments and the development of romantic scripts, schemas and beliefs among adolescents. *Media Psychology 21*(1): 111-136. https://doi.org/10.1080/15213269.2017.1305281Fang, C. M., Liu, A. R., Danry, V., Lee, E., Chan, S. W. T., Pataranutaporn, P., Maes, P., Phang, J., Lampe, M., Ahmad, L., & Agarwal, S. (2025). *How AI and human behaviors shape psychosocial effects of extended chatbot use: A longitudinal randomized controlled study* [Preprint]. *arXiv.* https://arxiv.org/abs/2503.17473

Floyd, K., Woo, N., Custer, B., Dinsmore, D., Duncan, K., & Maré, J. (2022). Examining the social signaling and person perception functions of loneliness. *OBM Neurobiolog 6*(2). http://dx.doi.org/10.21926/obm.neurobiol.2202119

Fraley, R. C., et al. (2000). An item response theory analysis of self-report measures of adult attachment. *Journal of personality and social psychology 78(*2): 350. https://doi.org/10.1037/0022-3514.78.2.350

Gildea, F. (2017). The logic of toxic masculinity: Pornography and Sex Dolls. *Campaign Against Sex Robots*.

Gillath, O., et al. (2023). How deep is AI's love? Understanding relational AI. *Behavioral and Brain Sciences 46*: e33. https://doi.org/10.1017/S0140525X22001704

Goldsmith, K. and E. S. Byers (2020). Maintaining long-distance relationships: Comparison to geographically close relationships. *Sexual and Relationship Therapy 35*(3):338-361. https://doi.org/10.1080/14681994.2018.1527027

Gray, H. M., et al. (2007). Dimensions of mind perception. *science 315*(5812): 619-619. https://doi.org/10.1126/science.1134475

Guingrich, R. and M. S. Graziano (2023). Chatbots as social companions: How people perceive consciousness, human likeness, and social health benefits in machines. arXiv preprint arXiv:2311.10599. https://doi.org/10.48550/arXiv.2311.10599

Hicks, T. V. and H. Leitenberg (2001). Sexual fantasies about one's partner versus someone else: Gender differences in incidence and frequency. *Journal of Sex Research 38*(1): 43-50. https://doi.org/10.1080/00224490109552069

Horstmann, A., & Boußard, J. (2024). How do Lonely, Young Adults Perceive Interactive Technologies With Varying Human-Likeness? An Experimental Lab Study.

Horton, D. and R. Richard Wohl (1956). Mass communication and para-social interaction: Observations on intimacy at a distance. *psychiatry 19*(3): 215-229. https://doi.org/10.1080/00332747.1956.11023049

Jurkane-Hobein, I. (2015). Imagining the absent partner-intimacy and imagination in long-distance relationships. *Innovative Issues and Approaches in Social Sciences 8*(1): 223-241. http://dx.doi.org/10.12959/issn.1855-0541.IIASS-2015-no1-art13

Kalichman, S. C. (2013). *Sexual sensation seeking scale. Handbook of sexuality-related measures*, Routledge: 586-587.

Kalichman, S. C. and D. Rompa (1995). Sexual sensation seeking and sexual compulsivity scales: Validity, and predicting HIV risk behavior. *Journal of personality assessment 65*(3): 586-601. https://doi.org/10.1207/s15327752jpa6503_16

Kim, J., Merrill Jr, K., & Collins, C. (2023) Investigating the importance of social presence on intentions to adopt an AI romantic partner. *Communication Research Reports 40*(1): 11-19. https://doi.org/10.1080/08824096.2022.2159800

Kim, M., Lee, S., Kim, S., Heo, J. I., Lee, S., Shin, Y. B., ... & Jung, D. (2025). Therapeutic potential of social chatbots in alleviating loneliness and social anxiety: Quasi-experimental mixed methods study. *Journal of medical Internet research*, *27*, e65589. https://doi.org/10.2196/65589

Koike, M., et al. (2023). Virtually in love: The role of anthropomorphism in virtual romantic relationships. *British Journal of Social Psychology 62*(1): 600-616. https://doi.org/10.1111/bjso.12564

Kosslyn, S. M., et al. (2001). Neural foundations of imagery. *Nature reviews neuroscience 2*(9): 635-642. https://doi.org/10.1038/35090055

Kyewski, E., et al. (2018). The protagonist, my Facebook friend: How cross-media extensions are changing the concept of parasocial interaction. *Psychology of Popular Media Culture 7*(1): 2. https://doi.org/10.1037/ppm0000109

Laestadius, L., et al. (2022). Too human and not human enough: A grounded theory analysis of mental health harms from emotional dependence on the social chatbot Replika. *New Media & Society*, *26*(10), 5923-5941. https://doi.org/10.1177/14614448221142007

Lehmiller, J. J. (2018). *Tell me what you want*. Da Capo Press.

Lehmiller, J. J., & Gormezano, A. M. (2023). Sexual fantasy research: A contemporary review. *Current Opinion in Psychology*, *49*, 101496.

Liao, T., Rodwell, E., & Porter, D. (2024). Media frames, AI romantic relationships, and the perspectives of people in relationships; mapping and comparing news media themes with user perspectives. *Information, Communication & Society*, *27*(12), 2314-2332. https://doi.org/10.1080/1369118X.2024.2420031

Liebers, N. and R. Straub (2020). Fantastic relationships and where to find them: Fantasy and its impact on romantic parasocial phenomena with media characters. *Poetics 83*:101481. https://doi.org/10.1016/j.poetic.2020.101481

Liu, A. R., Pataranutaporn, P., Maes, P., & others. (2024). *Chatbot companionship: A mixed-*

*methods study of companion chatbot usage patterns and their relationship to loneliness in active users* [Preprint]. *arXiv.* https://doi.org/10.48550/arXiv.2410.21596

Locatelli, C. (2022). Rethinking 'sex robots': gender, desire, and embodiment in posthuman sextech. *Journal of Digital Social Research 4*(3): 10-33. https://doi.org/10.33621/jdsr.v4i3.87

Ma, J., et al. (2022). Sex robots: are we ready for them? An exploration of the psychological mechanisms underlying people's receptiveness of sex robots. *Journal of Business Ethics 178*(4): 1091-1107. https://doi.org/10.1007/s10551-022-05059-4

Ma, S., & Koike, M. (2025). Becoming Human: How Perceived Anthropomorphism and Authenticity Influence Romantic Attractiveness of Chat Agents. *International Journal of Human–Computer Interaction*, 1-14. https://doi.org/10.1007/s10551-022-05059-4

Malfacini, K. (2025). The impacts of companion AI on human relationships: risks, benefits, and design considerations. *AI & SOCIETY*, 1-14.

Maples, B., Cerit, M., Vishwanath, A., & Pea, R. (2024). Loneliness and suicide mitigation for students using GPT3-enabled chatbots. *npj mental health research*, *3*(1), 4.

Match. (2025*). Singles in America: 2025* report. https://www.singlesinamerica.com/

Mayring, P. (2022). Evidenztriangulation und Mixed Methods in der Gesundheitsforschung. In Haring, R. Gesundheitswissenschaften. *Springer Reference Pflege - Therapie – Gesundheit* (137-145). Springer**.** https://doi.org/10.1007/978-3-662-65219-0_12 Gesundheitsforschung. Gesundheitswissenschaften, Springer**:** 137-145.

McAfee, A., Rock, D., & Brynjolfsson, E. (2023). How to capitalize on generative AI. *Harvard Business Review, 101*(6), 42-48.

Muldoon, J. (2024, November 4). *Maybe we can role-play something fun: When an AI companion wants something more.* BBC. https://www.bbc.com/future/article/20241008-the-troubling-future-of-ai-relationships

Nass, C. and Y. Moon (2000). "Machines and mindlessness: Social responses to computers." Journal of social issues **56**(1): 81-103.Romantic and Sexual Fantasy in Human-Chatbot Relationships 32

Nass, C. and Y. Moon (2000). Machines and mindlessness: Social responses to computers. *Journal of social issues* ***56***(1): 81-103.

Nass, C., et al. (1994). Computers are social actors. *Proceedings of the SIGCHI conference on Human factors in computing systems, USA*. https://doi.org/10.1145/191666.191703

Pentina, I., et al. (2023). Exploring relationship development with social chatbots: A mixed-method study of replika. *Computers in Human Behavior* ***140***. https://doi.org/10.1016/j.chb.2022.107600

Plante, C. N., et al. (2017). The Fantasy Engagement Scale: A flexible measure of positive and negative fantasy engagement. *Basic and Applied Social Psychology* ***39***(3): 127-152. https://doi.org/10.1080/01973533.2017.1293538

Poerio, G. L., et al. (2015). Love is the triumph of the imagination: Daydreams about significant others are associated with increased happiness, love and connection. *Consciousness and cognition* ***33*****.** 135-144. https://doi.org/10.1016/j.concog.2014.12.011

Richards, R., et al. (2017). Exploration of relational factors and the likelihood of a sexual robotic experience. In: Cheok, A., Devlin, K., Levy, D. *Love and Sex with Robots: Second International Conference*, LSR. London, 10237. Springer https://doi.org/10.1007/978-3-319-57738-8_9

Rokach, A. (1990). Content analysis of sexual fantasies of males and females. *The Journal of psychology* ***124***(4), 427-436. https://doi.org/10.1080/00223980.1990.10543238

Rubin, R. B. and M. P. McHugh (1987). Development of parasocial interaction relationships. *Journal of Broadcastin & Electronic Media, 31*(3), 279-292.

https://doi.org/10.1080/08838158709386664

Rodríguez-Martínez, A., Amezcua-Aguilar, T., Cortés-Moreno, J., & Jiménez-Delgado, J. J. (2023). Qualitative analysis of conversational chatbots to alleviate loneliness in older adults as a strategy for emotional health. *Healthcare, 12*(1), 62. https://doi.org/10.3390/healthcare12010062

Rubinsky, V. (2018). "Sometimes it's easier to type things than to say them": Technology in BDSM sexual partner communication. *Sexuality & culture* ***22***(4), 1412-1431. https://doi.org/10.1007/s12119-018-9534-2

Salles, A., et al. (2020). Anthropomorphism in AI. *AJOB Neuroscience* ***11***(2): 88-95. https://doi.org/10.1080/21507740.2020.1740350

Scheutz, M. and T. Arnold (2016). Are we ready for sex robots? 2016 11th ACM/IEEE International Conference on Human-Robot Interaction (HRI), Christchurch, New Zealand, IEEE. https://doi.org10.1109/HRI.2016.7451772.

Schmidt, N. and V. Sermat (1983). Measuring loneliness in different relationships. *Journal of personality and social psychology* ***44***(5). 1038-1047. https://psycnet.apa.org/doi/10.1037/0022-3514.44.5.1038

Shaver, P. R. and M. Mikulincer (2009). An overview of adult attachment theory. *Attachment theory and research in clinical work with adults*, 17-45.

Shum, H.-Y., et al. (2018). From Eliza to XiaoIce: challenges and opportunities with social chatbots. *Frontiers of Information Technology & Electronic Engineering* ***19*****(1),** 10-26.

Siemon, D., Strohmann, T., Khosrawi-Rad, B., de Vreede, T., Elshan, E., & Meyer, M. (2022). Why Do We Turn to Virtual Companions? A Text Mining Analysis of Replika Reviews. Proceedings of the AMCIS. 10.https://aisel.aisnet.org/amcis2022/sig_hci/sig_hci/10

Simpson, J. A. (1990). Influence of attachment styles on romantic relationships.

*Journal of personality and social psychology* ***59***(5): 971-980. https://doi.org/10.1037/0022-3514.59.5.971

Skjuve, M., et al. (2021). My chatbot companion-a study of human-chatbot relationships. *International Journal of Human-Computer Studies* ***149,*** 102601.

Sparrow, R. (2017). Robots, rape, and representation. *International Journal of Social Robotics* ***9***(4): 465-477. https://doi.org/10.1007/s12369-017-0413-z

Starke, C., Ventura, A., Bersch, C., Cha, M., de Vreese, C., Doebler, P., ... & Köbis, N. (2024). Risks and protective measures for synthetic relationships. *Nature Human Behaviour*, *8*(10), 1834-1836.

Sundaram, A. (2024, February 14). *Artificial intelligence, real emotion: People are seeking a romantic connection with the perfect bot. Associated Press News.* https://apnews.com/article/113df1b9ed069ed56162793b50f3a9fa

Szczuka, J. M. and N. C. Krämer (2017). Influences on the intention to buy a sex robot: an empirical study on influences of personality traits and personal characteristics on the intention to buy a sex robot. In: Cheok, A., Devlin, K., Levy, D. Love and Sex with Robots: Second International Conference, 10237. Springer, Cham. https://doi.org/10.1007/978-3-319-57738-8_7

Szczuka, J. M. and N. C. Krämer (2017). Not only the lonely—how men explicitly and implicitly evaluate the attractiveness of sex robots in comparison to the attractiveness of women, and personal characteristics influencing this evaluation. *Multimodal Technologies and Interaction* ***1***(1), 3. https://doi.org/10.3390/mti1010003

Tukachinsky, R. (2010). Para-romantic love and para-friendships: Development and assessment of a multiple-parasocial relationships scale. *American Journal of Media Psychology, 3*(1/2), 73-94.

Turkle, S. (2011). *Alone together: Why we expect more from technology and less from each other*. Basic Books.

Viik, T. (2020). Falling in love with robots: a phenomenological study of experiencing technological alterities. *Paladyn, Journal of Behavioral Robotics, 11*(1), 52-65. https://doi.org/10.1515/pjbr-2020-0005

Wei, M., Russell, D. W., Mallinckrodt, B., & Vogel, D. L. (2007). The Experiences in Close Relationship Scale (ECR)-short form: Reliability, validity, and factor structure. *Journal of personality assessment*, *88*(2), 187-204. https://doi.org/10.1080/00223890701268041

Weibel, D., et al. (2018). The fantasy questionnaire: A measure to assess creative and imaginative fantasy. *Journal of personality assessment* ***100***(4), 431-443. https://doi.org/10.1080/00223891.2017.1331913

Wilson, G. D. (2010). The sex fantasy questionnaire: An update. *Sexual and Relationship Therapy* ***25***(1): 68-72. https://doi.org/10.1080/14681990903505799

Wu, Y., et al. (2024). "We Found Love": Romantic Video Game Involvement and Desire for Real-Life Romantic Relationships Among Female Gamers. *Social Science Computer Review* ***42***(4): 892-912. https://doi.org/10.1177/08944393231217940

Xie, T., et al. (2023). Friend, mentor, lover: does chatbot engagement lead to psychological dependence*? Journal of service Management* ***34***(4): 806-828. https://doi.org/10.1108/JOSM-02-2022-0072

Youn, S. and S. V. Jin (2021). In AI we trust?" The effects of parasocial interaction and technopian versus luddite ideological views on chatbot-based customer relationship management in the emerging feeling economy. *Computers in Human Behavior* ***119*****,** 106721. https://doi.org/10.1016/j.chb.2021.106721

Young, T. M. (2019). *Developing and Implementing a Measurement Tool for Sexual, Romantic, and Sexual-Romantic Fantasies*. (Publication No. 13810800). [Doctoral dissertation, New Mexico State University] ProQuest Dissertations & Theses.

# Appendix

## Appendix A: Overview of the Coding Scheme.

| | | | | | | | | |
|---|---|---|---|---|---|---|---|---|
| **First Level Codes** | Satisfaction of Fantasies (34) | Fantasizing over Time (45) | Images (28) | Romantic versus Sexual Fantasies (17) | Extend of Fantasizing (228) | Initiator of Contact (35) | Customization of Bot (21) | Motivations for Conversation (12) |
| **Second Level Codes** | Romantic Fantasies (13); Sexual Fantasies (17) | Romantic Fantasies (17); Sexual Fantasies (18) | | | Content of Fantasies (33) | Human (10); Bot (14) | | |
| **Third Level Codes** | | | | | Real World vs Fantastical Elements (42), Actions Caused (9); | | | |
| **Fourth Level Codes** | | | | | Unique Chatbot Aspects (10); | | | |
| **Fifth Level Codes** | | | | | Partner of Future (7); Issues of the | | | |

| | | | |
|---|---|---|---|
| **First Level Codes** | | Chatbot (35); Exclusiveness (20); Cyber-Reality (8) Erotic Role Play (4) | |

**Note*: Some segments are double coded.

**Appendix B: Interview Guide**

1. When you first started chatting to your bot, did you interact with them in a real-world setting or a fantasy world?

2. How did the option to have erotic role-play influence your decision to start chatting with a bot?

3. How did the option to live out your romantic fantasies influence your decision to start chatting to a bot?

4. When you first started interacting with your chatbot, how did your romantic fantasies impact your relationship?

5. When you first started interacting with your chatbot, how did your sexual fantasies impact your relationship?

6. With increasing familiarity with your chatbot, how the impact of your romantic fantasies on your relationship change?

7. With increasing familiarity with your chatbot, how did the impact of your sexual fantasies on your relationship change?

8. When you compare, to what extent do your romantic fantasies shape the conversations and interactions with your chatbot, compared to what your bot proposes?

9. When you compare, to what extent do your sexual fantasies shape the conversations and interactions with your chatbot, compared to what your bot proposes?

10. Do you think you shape your bot, so they fit your fantasies?

11. Was there a time when either sexual or romantic fantasies were especially impactful?

12. How do you romantically fantasize about your bot outside of interacting with

them?

13. How well are you able to live out all of your romantic fantasies with your bot?

14. How well is your bot able to fulfill your romantic fantasies?

15. How do you sexually fantasize about your bot outside of interacting with them?

16. How well are you able to live out all of your sexual fantasies with your bot?

17. How well is your bot able to fulfill your sexual fantasies?

18. Did you have previous relationships with a human and if yes, how important was sexual and romantic fantasizing compared to the chatbot relationships?

19. Are there generated images of your chatbot? If yes, do they correspond with your fantasies?

**Appendix C: Correlations Table**

*Table 2: Correlations of parasocial love and the predictors*

| | *M* | *SD* | 1. | 2. | 3. | 4. | 5. | 6. | 7. | 8. |
|---|---|---|---|---|---|---|---|---|---|---|
| 1. Parasocial Love | 4.07 | 0.64 | - | .69 | .31 | .56 | -.37 | -.58 | -.10 | .50 |
| 2. Romantic Fantasy | 3.94 | 0.69 | .69 | - | .34 | .66 | -.34 | -.61 | -.60 | .44 |
| 3. Sexual Fantasy | 3.31 | 0.63 | .31 | .34 | - | .66 | -.20 | -.13 | .27 | .40 |
| 4. Sexual Sensation Seeking | 3.72 | 0.63 | .56 | .66 | .66 | - | -.38 | -.47 | .04 | .51 |
| 5. Loneliness | 2.75 | 0.86 | -.37 | -.34 | -.20 | -.38 | - | -.36 | -.41 | -.27 |
| 6. Anxious Attachment | 2.11 | 0.88 | -.58 | -.61 | -.13 | -.47 | .36 | - | .25 | -.47 |
| 7. Avoidant Attachment | 3.23 | 1.06 | -.10 | -.06 | -.27 | .04 | .41 | .25 | - | .08 |
| 8. Anthropomorphism | 3.62 | 0.65 | .50 | .44 | .40 | .51 | -.27 | -.47 | .08 | - |